\documentclass[aps,prl,twocolumn,showpacs,psfig]{revtex4}
\input epsf.tex
\usepackage{psfig}
\include{bibliohead}

\def\zabs{$z_{\rm abs}$}

\def\h2{H$_2$}

\def\kms{km~s$^{-1}$}
\def\dela{$\Delta\alpha/\alpha$~}
\begin{document}

%
\preprint{}
\title{Limits on the time variation of the electromagnetic
    fine-structure constant in the low energy limit from absorption
    lines in the spectra of distant quasars}
\author{R. Srianand$^1$, H. Chand$^1$, P. Petitjean$^{2,3}$ and B. Aracil$^2$ }
%
\affiliation{$^1$IUCAA, Post Bag 4, Ganeshkhind, Pune - 411 007\\
$^2$Institut d'Astrophysique de Paris -- CNRS, 98bis Boulevard Arago, F-75014 
Paris, France\\
$^3$LERMA, Observatoire de Paris, 61 avenue de l'Observatoire, F-75014 Paris, 
France}
\email{anand@iucaa.ernet.in, hcverma@iucaa.ernet.in, ppetitje@iap.fr, 
aracil@iap.fr}
\date{\today}
\begin{abstract}
Most of the successful physical theories rely on the constancy of
few fundamental quantities (such as the speed of light, $c$, the 
fine-structure constant, $\alpha$,  the proton to electron mass ratio,
$\mu$, etc), and constraining the possible time variations of these 
fundamental quantities is an important step toward 
a complete physical theory.
Time variation of $\alpha$ can be accurately probed using 
absorption lines seen in the spectra of distant quasars.
Here, we present the results of a detailed many-multiplet 
analysis performed on a new sample of Mg~{\sc ii} systems 
observed in high quality quasar spectra obtained using 
the Very Large Telescope.
The weighted mean value of the variation in ${\bf \alpha}$ derived from 
our analysis over the redshift range ${\bf 0.4\le z\le 2.3}$ is 
${\bf \Delta\alpha/\alpha}$~=~${\bf (-0.06\pm0.06)\times10^{-5}}$. 
The median redshift of our sample (z$\simeq$1.55) corresponds to
a look-back time of 9.7 Gyr in the most favored cosmological model today.
This gives a 3$\sigma$ limit, ${\bf -2.5\times 10^{-16} ~{\rm yr}^{-1}\le(\Delta\alpha/\alpha\Delta t) 
\le+1.2\times 10^{-16}~{\rm yr}^{-1}}$, for the time variation of $\alpha$,
that forms the strongest constraint obtained based on high redshift quasar 
absorption line systems.
\end{abstract}
%
%
\maketitle

Contemporary theories of fundamental interactions allow the  
fundamental constants to vary as a function of space and
time\cite{uzan}. Therefore, direct measurements of possible
time variations of various fundamental quantities are of utmost 
importance for a complete understanding of fundamental physics.
In this letter we investigate the time variation of the 
fine-structure constant ($\alpha { =e^2/\hbar c}$~=~1/137.03599976(50)
{\cite{mohr}}, where $e$ is the charge of the electron and $\hbar$ the 
reduced Planck constant  measured in the laboratories on Earth). 
The time evolution of $\alpha$ can be probed in the framework of 
standard Big-Bang models using measurements performed at different
redshifts ($z$). The Oklo phenomenon\cite{fujii}, a natural fission 
reactor that operated 2~Gyrs ago, or at $z$~$\sim$~0.16, gives the strongest 
constraint on the variation of $\alpha$ over cosmological time-scales,
$[\Delta\alpha/\alpha\Delta t]=(-0.2\pm0.8)\times 10^{-17}$~yr$^{-1}$.
At higher redshifts, a possible time dependence will be registered 
in the form of small shifts in the absorption line spectra seen toward 
distant quasars\cite{save} as the energy of the atomic 
transitions depend on $\alpha$.
Initial attempts to measure the variation of
$\alpha$ were based on the absorption lines of 
alkali-doublets\cite{wolfe}(AD method). The best constraint 
obtained using 
this method is $\Delta\alpha/\alpha$ = $(-0.5\pm1.3)\times10^{-5}$\cite{mur1}.
Other methods such as the one using O~{\sc iii} emission lines\cite{bah1}, 
though more robust, are not sensitive enough to detect small variations 
in \dela. Investigations based on molecular lines\cite{murmol} 
detected in two systems give
\dela = $(-0.10\pm0.22)\times10^{-5}$ and $(-0.08\pm0.27)\times
10^{-5}$ at \zabs = 0.2467 and 0.6847 respectively. Such studies
at high-z are elusive due to lack of molecular systems.
The generalization of alkali-doublet method, called many-multiplet 
(MM) method, gives an order of magnitude improvement in the measurement of 
\dela compared to AD method\cite{dzu1} by using not only
doublets but several multiplets from different species. The sensitivity of
different line transitions from different multiplets 
to variations in $\alpha$ were
computed using many-body calculations taking into account
dominant relativistic effects\cite{dzu2}.
\par
In simple terms, MM method exploits the fact that the energy of 
different line transitions vary differently for a given change 
in $\alpha$. For example rest wavelengths of
Mg~{\sc ii}$\lambda\lambda$2797,2803 and  Mg~{\sc i}$\lambda$2852 
transitions are fairly insensitive to small changes in 
$\alpha$ thereby providing good anchor for measuring the 
systemic redshift. Whereas the rest wavelengths 
of Fe~{\sc ii} multiplets are quite sensitive to small variations 
in $\alpha$. Thus measuring consistent relative shifts between an 
anchor and different Fe~{\sc ii} lines can in principle lead to an
accurate measurement of $\Delta\alpha/\alpha$. 
The accuracy at which the variation can be 
measured depends very much on how well absorption profiles
can be modeled. For this usually Voigt profiles that are convolved 
with the instrumental profile and characterized 
by column density ($N$), velocity dispersion ($b$)  and redshift 
in addition to the rest-wavelength of the species
are used.
In a real spectrum small relative shifts can be introduced 
by various systematic effects such as inhomogeneities in the 
absorbing region, poor wavelength calibration, isotopic abundances, 
and atmospheric dispersion effects, etc\cite{mur01}. 
However most of the random systematic 
effects can be canceled by using a large number of measurements. 
MM method applied to large samples of quasar absorption lines resulted
in the claim for smaller value of $\alpha$ in the 
past\cite{dzu1}, $\Delta\alpha/\alpha$ = $(-0.574\pm0.102)\times10^{-5}$
for 0.2$\le z\le$3.7. 
The main motivation of the present work is to
perform similar analysis using a completely different line
fitting code and an independent, uniform,
better quality, and well defined data sample.
\par
The data used in this study were obtained with the Ultra-violet and Visible 
Echelle Spectrograph (UVES) mounted on the ESO Kueyen 8.2~m telescope 
at the Paranal observatory for the ESO-VLT  Large Programme 
``QSO absorption lines". This programme has been devised to
gather a homogeneous sample of echelle spectra of 18 quasars, 
with the uniform spectral coverage,  resolution and  signal-to-noise
ratio (S/N) suitable for studying the intergalactic medium 
in the redshift range 1.7$-$4.5. 
Spectra of quasars were obtained in 
service mode observations spread over 4 periods
(2 years) covering 30 nights under good seeing conditions ($\le0.8$ arcsec)
using a slit 1 arcsec wide. 
The data were reduced using the
UVES pipeline, a standard set of procedures implemented in a dedicated context 
of MIDAS, the ESO data reduction package. However independent consistency
checks were made to ensure good sky-background subtraction
(better than 1\% accuracy), accurate
wavelength calibration (rms in $\Delta\lambda$/$\lambda$
better than $\sim$~4$\times$10$^{-7}$ over the full wavelength range
of interest, 3100$-$5400 and 5450$-$9000~\AA)
and optimum signal to noize (S/N~$\sim$~50$-$80 per pixel).
We fit the 
continuum using lower order polynomial that gives $\chi^2\simeq 1$. 
\begin{figure}[]
\centerline{\vbox{
\psfig{figure=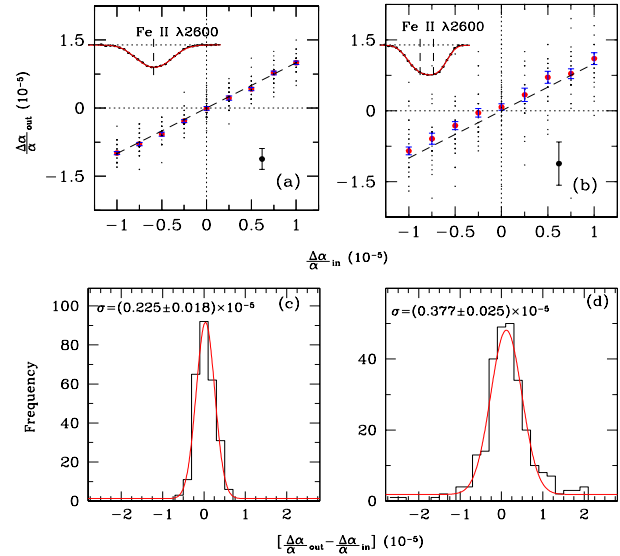,height=8.cm,width=8.5cm,angle=0.}
}}
\caption[]{
Absorption spectra of Mg~{\sc ii} and Fe~{\sc ii} were 
simulated for a given set of $N$, $b$, and spectral resolution 
similar to our data, introducing spectral shifts corresponding to a given
value of \dela. Top panels show the relationship between
the input and derived value of $\Delta\alpha/\alpha$ in the case
of a single clean component (left-hand side) and a blend of two components 
(right-hand side). Each realization is performed using random values
of $N$, $b$ and noise keeping the signal-to-noise ratio, wavelength 
sampling and resolution as in a typical UVES spectrum. 
Dots are the values from individual 
realizations and the points with the error bars are the weighted mean 
obtained from 30 realizations. A typical absorption profile is also shown in these panels. 
The lower panels give the distribution of the recovered $\Delta\alpha/\alpha$
around the true one. Single (left) and blended (right) cases are
considered respectively. 
Best fitted Gaussian distributions are over-plotted.
}
\label{fig1}
\end{figure}
\begin{figure}[]
\centerline{\vbox{
\psfig{figure=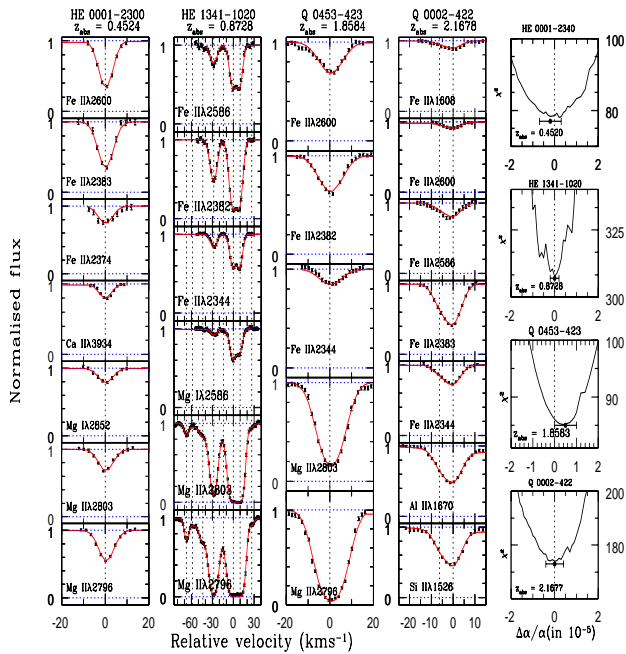,height=9.cm,width=8.5cm,angle=270.}
}}
\caption[]{
Voigt profile fits to 4 randomly chosen  systems 
(out of 23) in our sample are shown
in the first 4 columns from the left. The quasar name and absorption
redshifts are given on top of the panels. The points with
the error bars are the observed data and the continuous
curve is the fit obtained using multicomponent Voigt
profile decomposition. The locations of different 
components are marked with vertical dotted lines. 
The plots in the right most column demonstrate how \dela is extracted 
from these systems. We plot $\chi^2$ as the function of
\dela. The minimum in this curve (marked with a dot)
gives the best-fit value of \dela and the error in
this measurement (error-bar around the dot) is based
on the standard statistical method of computing errors
from $\Delta\chi^2 = 1$.
} 
\label{fig2}
\end{figure}
Air-vacuum conversions and heliocentric corrections were done using
standard conversion equations\cite{edlen}. The slit was always oriented 
along the parallactic angle and calibration exposures were taken
before or after the scientific exposures. The full
descriptions of the data reduction and calibration procedures, 
sample selection and all the steps involved in the analysis 
are presented in Chand et al.,(2004)\cite{chand}. 
\par
As a first step we have performed a detailed analysis on simulated data to have
a clearer understanding of various possible systematics that can affect the
analysis of real data. The results of this exercise are used
to  validate our procedure and define selection criteria 
that will minimise systematics in our analysis.
Absorption spectra of Mg~{\sc ii} and Fe~{\sc ii} were
simulated for given $N$, and $b$. The spectral resolution and S/N ratio 
were similar to our UVES data.
The standard fitting function and coefficients\cite{dzu2} are used
to incorporate the effects of $\alpha$ variation.
We considered two cases: 
a simple single component system and a highly blended two-component system.
In the highly blended case we restricted the separation between 
the two components to be always smaller than the velocity 
dispersion of one of the components. We then fitted the absorption lines 
in order to recover \dela introduced in the input spectrum.
This exercise is to determine the precision that can be reached using 
our method and fitting procedure. 
For this we used the Voigt profile fitting method and
standard $\chi^2$ statistics to fit the absorption profiles consistently 
and to determine the best fit value for \dela. The relationships between the
input and recovered values of \dela are shown in Fig.~\ref{fig1}.
The method works very well in the case
of simple single component systems where one expects minimum
uncertainties due to systematics. The deviation of the
recovered value with respect to the true one distributes
like a Gaussian with $\sigma = 0.23\times 10^{-5}$. 
This shows that \dela can be constrained with an accuracy of 
0.07$\times 10^{-5}$ even when we use 10 such systems.
As expected the scatter is more in the case of 
blended systems ($\sigma = 0.38\times 10^{-5}$). 
Similar analysis on two components with separations larger than
individual $b$ values gives $\sigma = 0.261\times10^{-5}$.
Indeed, in this case, the separation of the components is large 
enough so that each component is clearly recognized by the 
fitting procedure. In such a situation the 
corresponding errors are very much the same as for individual
components. The conclusion is that it is better to select
single component or not heavily blended systems,  
that is the components are separated by more than their $b$ values,
to obtain the most reliable results. We notice from our simulations 
that systems with weak lines have large uncertainties in the measured 
\dela and should be avoided from the analysis. Finally, the data 
used in this analysis have a median S/N ratio of 70 per pixel, a 
factor of two better than that used in earlier studies.
Based on the simulations we expect this enhancement in the S/N
ratio to provide a roughly a factor two improvement in the \dela 
measurements.
\par
In summary, based on the results from the simulations, we apply the following 
selection criteria to derive reliable \dela:
(i) we consider only species with  similar ionization potentials 
(Mg~{\sc ii}, Fe~{\sc ii}, Si~{\sc ii} and Al~{\sc ii}) as they are 
most likely to originate from similar regions in the cloud;
(ii) we avoid absorption lines that are contaminated by atmospheric lines;
(iii) we consider only systems that have 
$N$(Fe~{\sc ii})$\ge 2\times10^{12}$ cm$^{-2}$ which ensures
that all the standard Fe~{\sc ii} multiplets are detected 
at more than  5$\sigma$ level;
(iv) we demand that at least one of the anchor lines is
not saturated so that the redshift measurement is robust;
(v) we also avoided sub-DLAs (i.e $N$(H~{\sc i})$\ge10^{19}$ cm$^{-2}$) 
as these systems may have ionization and chemical
inhomogeneities;
(vi) we do not consider strongly saturated systems with large velocity spread
(complex blends); however in such systems
whenever we find a well detached satellite components we 
include these components in the analysis;
(vii) finally, based on the component structure resulting from
the Voigt profile fits of systems that are not complex
blends, we retain only 
systems for which the majority of components are separated
from its neighboring components by more than the $b$ parameters. 
Application of the above conditions resulted in 23 systems
for performing \dela measurements(6 single component systems, 
6 well separated doubles, 6 systems with 3 components 
with at least one well detached from the rest, and  5 
systems with more than 3 components) .
\par
In our analysis we use the fact that relative shifts between lines from the 
same species (say Fe~{\sc ii} multiplets) are insensitive
to \dela. We first fit all the Fe~{\sc ii} lines (except for Fe~{\sc ii}$\lambda$1608) simultaneously using 
laboratory wavelengths. This allows us to find out about (i) bad pixels, 
(ii) unknown contaminations and (iii) the velocity component structure 
in Fe~{\sc ii}. Defective absorption lines detected are removed from 
the \dela analysis. Similar exercise was carried out for Mg~{\sc ii} 
doublets and other anchors.  Based on these preliminary fits a 
first set of parameters is generated to start the Voigt profile 
fitting procedure that includes \dela variations.
We illustrate the method in Fig.~\ref{fig2} using 4 randomly chosen
systems from our sample. 
We vary \dela  ranging from $-5.0\times10^{-5}$ to $5.0\times10^{-5}$ 
in step of $0.1\times10^{-5}$, and each time fit all the 
lines together. $\chi^2$ minima obtained for each of these fits
are plotted as a function of \dela (right most panel in Fig.~{\ref{fig2}}).
The value of \dela at which this $\chi^2$ is minimum is accepted as
the measure of best possible \dela value. Following standard statistical
procedure we assign 1$\sigma$ error bar to the best fitted
value of \dela by computing the change in 
\dela implying $\Delta\chi^{2}= \chi^{2}-\chi^2_{min}=1 $. 
We always consider two different models:
(i) one in which $b$ parameters for all the species are the same and (ii) 
one in which $b$ parameters for
different species are different. In all systems we notice that the 
derived \dela using both cases are consistent with one another
within $1\sigma$ uncertainty. We use the result that has a lower 
reduced $\chi^2$ for our final analysis. 
\par\noindent
\begin{figure}
\centerline{\vbox{
\psfig{figure=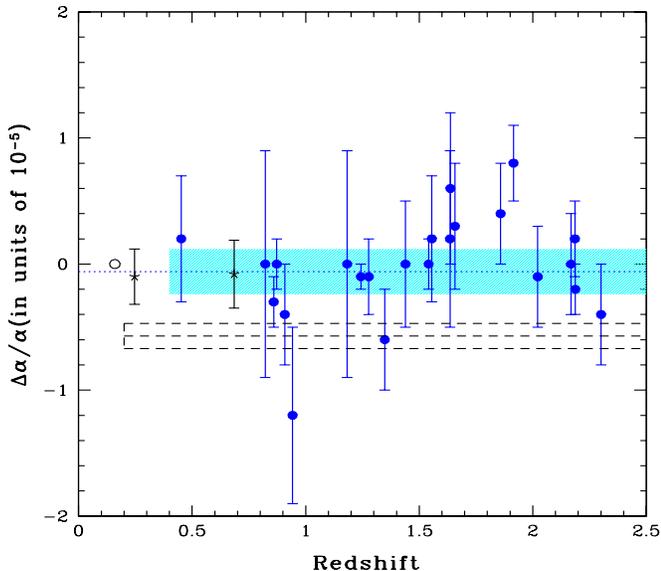,height=8.cm,width=9.cm,angle=0.}
}}
\caption[]{ 
The measured values of $\Delta\alpha/\alpha$ from
our sample (filled circles) are plotted
against the absorption redshifts of Mg~{\sc ii} systems. Each point is the 
best fitted value obtained for individual systems using $\chi^2$ 
minimization as demonstrated in Fig.~\ref{fig2}. 
The open circle and stars are the measurement from 
Oklo phenomenon{\cite{fujii}} and from molecular lines{\cite{murmol}}
respectively. The weighted mean and 1$\sigma$ range measured by 
Murphy et al.(2003) are shown with the horizontal long dashed lines. 
Clearly most of our measurements are inconsistent with this range.
The shadow region marks the weighted mean and its 3$\sigma$ error 
obtained from our study.
}
\label{fig3}
\end{figure}
Results obtained for the 23 systems in our sample and
that from the literature are summarized in  Fig.~\ref{fig3}.
The shaded region passing through most of the error bars is 
the weighted mean (with 1/error$^2$ weights)
and its 3$\sigma$ error from our sample.
%
It is clear that most of our measurements are 
consistent with zero within the uncertainties.
The simple mean, weighted mean and standard deviation
around the mean obtained for our sample 
are $(-0.02\pm0.10)\times10^{-5}$, $(-0.06\pm0.06)\times10^{-5}$
and 0.41$\times10^{-5}$ respectively. The corresponding values 
are $(0.01\pm0.15)\times10^{-5}$,  $(-0.08\pm0.07)\times10^{-5}$
and 0.26$\times10^{-5}$ respectively when we restrict our analysis to 
the single and well detached two component systems (12 in total).
These systems have maximum influence in the weighted mean of the 
whole sample. Interestingly, error on the weighted mean and 
standard deviation around the mean  from these systems are very 
close to the predictions from our simulations. In all cases
weighted mean values are consistent with all the data points with a 
reduced $\chi^2\simeq1$. 
\par
Thus our study gives a more stringent $3\sigma$ constraint 
of $-0.24\le \Delta\alpha/\alpha ({\rm ~in~10^{-5}})  \le +0.12$ 
over the redshift range of 0.4$\le z\le$2.3.
The median redshift of the whole sample is 1.55 which corresponds
to a look-back time of 9.7 Gyr
in the most favored cosmological model today
(${\rm \Omega(total) = 1}$, ${\rm \Omega_\Lambda = 0.7}$, 
${\rm \Omega(matter) = 0.3}$, and H$_0$ = 68 \kms Mpc$^{-1}$).
This gives a 3$\sigma$ constraint on the time variation of
\dela to be 
$-2.5\times 10^{-16} ~{\rm yr}^{-1}\le(\Delta\alpha/\alpha\Delta t) 
\le 1.2\times 10^{-16}~{\rm yr}^{-1} $. 
As can be seen from Fig.~\ref{fig3} our results are consistent with
the constraints from Oklo phenomenon and from the molecular lines.
However, our study does not support the claims by previous authors
of a statistically significant change in \dela with 
cosmic time at $z$~$>$~0.5 using MM method. In any case
our measurement still does allow smaller variations in 
excess of what is found based on the Oklo phenomenon. Future very 
high resolution (R$\sim$100,000) spectroscopic
studies are needed to probe the variations in $\alpha$ with much 
better accuracy.
\acknowledgments
This work is based on observations collected during programme 166.A-0106 
(PI: Jacqueline Bergeron) of the European Southern Observatory with the 
Ultra-violet and Visible Echelle 
Spectrograph mounted on the 8.2~m Kueyen telescope operated at the Paranal 
Observatory, Chile. HC thanks CSIR, INDIA for the grant award
No. 9/545(18)/2KI/EMR-I. RS thanks CNRS/IAP
for the hospitality.
We gratefully acknowledge support from the Indo-French 
Centre for the Promotion of Advanced Research (Centre Franco-Indien pour 
la Promotion de la Recherche Avanc\'ee) under contract No. 3004-A. 

\end{document}